\documentstyle[12pt]{article}
\newcommand{\be}{\begin{equation}}
\newcommand{\ee}{\end{equation}}
\newcommand{\bea}{\begin{eqnarray}}
\newcommand{\ena}{\end{eqnarray}}
\newcommand{\eea}{\end{eqnarray}}
\newcommand{\beano}{\begin{eqnarray*}}
\newcommand{\enano}{\end{eqnarray*}}

\newcommand{\ca}{\mbox{$\cal{A}$}}

\newcommand{{\cf}}{\mbox{$\cal{F}$}}
\newcommand{{\cg}}{\mbox{$\cal{G}$}}

\newcommand{\cq}{\mbox{$\cal{Q}$}}

\newcommand{\pp}{{=\!\!\! |}}
\newcommand{\del}{\partial}
\newcommand{\bdel}{\bar{\partial}}

\def\vf{\varphi}
\def\a{\alpha}
\def\b{\beta}

\def\d{\delta}

\def\g{\gamma}
\def\h{\eta}

\def\j{\psi}
\def\k{\kappa}

\def\m{\mu}

\def\p{\pi}

\def\t{\tau}

\def\P{\Pi}

\newcommand{\nonu}{\nonumber \\[2mm]}

\begin{document}
\newcommand{\eqn}[1]{eq.(\ref{#1})}

\renewcommand{\section}[1]{\addtocounter{section}{1}
\vspace{5mm} \par \noindent
  {\bf \thesection . #1}\setcounter{subsection}{0}
  \par
   \vspace{2mm} } 
\newcommand{\sectionsub}[1]{\addtocounter{section}{1}
\vspace{5mm} \par \noindent
  {\bf \thesection . #1}\setcounter{subsection}{0}\par}
\renewcommand{\subsection}[1]{\addtocounter{subsection}{1}
\vspace{2.5mm}\par\noindent {\em \thesubsection . #1}\par
 \vspace{0.5mm} }
\renewcommand{\thebibliography}[1]{ {\vspace{5mm}\par \noindent{\bf
References}\par \vspace{2mm}}
\list
 {\arabic{enumi}.}{\settowidth\labelwidth{[#1]}\leftmargin\labelwidth
 \advance\leftmargin\labelsep\addtolength{\topsep}{-4em}
 \usecounter{enumi}}
 \def\newblock{\hskip .11em plus .33em minus .07em}
 \sloppy\clubpenalty4000\widowpenalty4000
 \sfcode`\.=1000\relax \setlength{\itemsep}{-0.4em} }
\hfill\hbox{VUB-TH.495, hep-th/9511050}\\[1.4cm]
\vspace{4mm}
\begin{center}
{\bf GAUGING WESS-ZUMINO-WITTEN MODELS }\footnote{Contribution to the
proceedings of the {\it Workshop on gauge theories, applied supersymmetry and
quantum gravity}, Leuven, Belgium, july 1995}

\vspace{1.4cm}

ALEXANDER SEVRIN \\
{\em Theoretische Natuurkunde, Vrije Universiteit Brussel} \\
{\em Pleinlaan 2, B-1050 Brussel, Belgium} \\
\end{center}
\centerline{ABSTRACT}
\vspace{- 4 mm}  
\begin{quote}\small
We review some aspects of gauged WZW models. By choosing a solvable
subgroup as gauge group, one is lead to three main applications: the
construction of field theories with an extended conformal symmetry, the
construction of the effective action of (extended) 2D gravities and the
systematic construction of string theories with some extended gauge
symmetry.
\end{quote}
\addtocounter{section}{1}
\par \noindent
  {\bf \thesection . Introduction}
  \par
   \vspace{2mm} 
\noindent
For any semi-simple (super) Lie algebra ${\cal G}$, one can write down the
corresponding
Wess-Zumino-Witten model \cite{witten}. Its action is given by ${\cal
S}=\kappa{\cal
S}^+$ with $\kappa$ the level and ${\cal S}^+$ given by
\begin{eqnarray}
S^+ [g]= \frac{1}{4\pi x} \int d^2 z \; str \left\{ \partial g^{-1}
\bar{\partial} g
\right\} + \frac{1}{12\pi x}
\int d^3 z\; \varepsilon^{\alpha \beta\gamma} \, str \left\{ g_{,\alpha }
g^{-1} g_{,\beta}
g^{-1} g_{,\gamma} g^{-1} \right\},
\label{nineteen}
\end{eqnarray}
with $x$ the index of the representation.
The functional   ${\cal S}^+$ satisfies the Polyakov-Wiegman identity
\cite{pw}:
\begin{eqnarray}
S^+[hg]=S^+[h]+S^+[g]-\frac{1}{2\pi x}\int str\Bigl( h^{-1}\partial
h\bar\partial g
g^{-1} \Bigr) .\label{pwfor}
\end{eqnarray}
We will also use the functional $S^-[g]$, defined by
$S^-[g]=S^+[g^{-1}]$.
Using the equations of motion of the WZW model,
\begin{eqnarray}
\delta S^+[g]=\frac{1}{2\pi x}\int str\left\{\bar\partial (g^{-1}\partial
g)g^{-1}\delta
g\right\}
=\frac{1}{2\pi x}\int str\left\{\partial(\bar\partial g g^{-1})\delta g
g^{-1}\right\},
\end{eqnarray}
one gets
the conserved affine currents
\begin{eqnarray}
J_z=-\frac{\kappa}{2}g^{-1}\partial g,\qquad
J_{\bar z}=\frac{\kappa}{2}\bar\partial g g^{-1}.
\end{eqnarray}
The OPE's of the affine currents close:
\begin{eqnarray}
J^a_{z} (z) J^b_{z} (w) = - \frac{\kappa}{2} g^{ab} (z-w)^{-2} + (z-w)^{-1}
f^{abc}
J_{z\, c} (w) + \cdots,
\end{eqnarray}
where we defined the Killing metric by $str(t_a t_b)=-x\,g_{ab}$. A
similar relation holds for $J_{\bar z}$. The energy-momentum
tensor of the WZW model is given by the Sugawara construction
\begin{eqnarray}
T=\frac{1}{x\left(\kappa+\tilde{h} \right)}str J_zJ_z,
\end{eqnarray}
with $\tilde{h}$ the dual Coxeter number
and it satisfies the Virasoro algebra with the central extension given by:
\begin{eqnarray}
c=\frac{\kappa (d_B-d_F)}{\kappa+\tilde{h}},
\end{eqnarray}
where $d_B$, $d_F$ resp., is the number of bosonic, fermionic resp.,
generators of the (super) Lie algebra.

Not only does the WZW model provide us with a large class of non-trivial
conformal field theories (cft's), it can also be used to construct other cft's.
The way to achieve this goes through gauging the WZW model.
Using the Polyakov-Wiegman identity, it is straightforward to
gauge any subalgebra of ${\cal G}$. Indeed, $\kappa S^-[h_+gh_-]$,
with $h_\pm\in{\cal H}_\pm\subseteq{\cal G}$, can be worked out using eq.
(\ref{pwfor}) and
the result is clearly invariant under gauge transformations
\begin{eqnarray}
g\rightarrow \gamma^{-1}_+g\gamma^{-1}_-,\qquad h_+\rightarrow
h_+\gamma_+,\qquad
h_-\rightarrow \gamma_- h_-.
\end{eqnarray}
However, if one tries to write out the action in terms of the gauge fields
$A_z=\partial h_-h_-^{-1}$ and $A_{\bar z}=-h_+^{-1}\bar\partial h_+$, one
generically obtains a non-local expression. A
well-known way out is taking $ {\cal H}_+=  {\cal H}_-$ and choosing $\kappa
(S^-[h_+gh_-]-
S^-[h_+h_-])$ as the gauge invariant action. Another possibility is choosing
the
subgroup to be such that all non-local terms vanish. Precisely the last
case will be studied in this paper.
\setcounter{equation}{0}
\section{sl(2) embeddings and extended conformal symmetries}
\noindent
We consider a non-trivial embedding of $sl(2)$ into a (super) Lie algebra
${\cal G}$.
The adjoint representation  of ${\cal G}$ decomposes into $sl(2)$ irreps:
\bea
\mbox{adjoint}(\cg)=\bigoplus_{j\in\frac 1 2 {\bf N} }n_{j}\,
\underline{2j+1},\label{decomp}
\eea
with $n_{j}\in{\bf N}$, the multiplicities. The $sl(2)$ embedding induces
a natural grading on ${\cal G}$ given\footnote{The $sl(2)$ generators are
denoted by $\{e_\pp,e_=, e_0\}$ and they satisfy $[e_0,e_\pp]=2e_\pp\,$,
$[e_0,e_=]=-2e_=$ and
$[e_+, e_-]=e_0$.} by the eigenvalues of $e_0$. In \cite{bais}, it was
shown that constraining the affine currents of the coresponding WZW model
as:
\begin{eqnarray}
J=T+\frac\kappa 2 e_=,\label{bbais}
\end{eqnarray}
where $J$ stands for $J_z$ and $T\in\ker\,\mbox{ad}_{e_{\pp}}\cg$,
breaks the affine symmetry down to some extension of the Virasoro
symmetry which is generated by $n_j$ currents of conformal dimension
$j+1$ for $j\in\frac 1 2 {\bf N}$. In order to realize this in terms of an
action, we consider a WZW model on ${\cal G}$ where we gauge the
subalgebra $\Pi_{>0}{\cal G}$. We use the notation that for $X\in{\cal G}$,
$\Pi_{>0}X$ projects out the strict positively graded part of $X$, with
the grading given by the $sl(2)$ induced grading. The gauge invariant
action is
\bea
{\cal S}=\k S^-[g]+\frac{1}{\pi x}\int str\, A\,J,
\eea
where $A$ stands for $A_{\bar z}$ and $A\in\P_{>0}{\cal G}$.
However, this action has no chance of reproducing eq. (\ref{bbais}), as
the equation of motion for $A$ puts $\Pi_{<0}J$ to zero. Instead, we take
\cite{wt}:
\bea
{\cal S}=\k S^-[g]+\frac{1}{\pi x}\int str\, A(J-\frac \k 2 e_=-\frac \k 2
[e_=,{\tau}])
-\frac{\k}{4\p x}\int str [e_=,{\tau}] \bdel{\tau},\label{actionst}
\eea
where $\t\in\Pi_{1/2}{\cal G}$ was introduced in order to obtain a gauge
invariant action. One checks that eq. (\ref{actionst}) is
invariant under:
\bea
\d g= \h g,\qquad
\d A= \bdel \h + {[}\h, A{]},\qquad
\d {\tau}= - \Pi_{\frac 1 2}\h,\label{trsfst}
\eea
and $\h\in \Pi_{>0}\cg$. One can verify that a combination of the equation
of motion of $A$ and a gauge choice, precisely reproduces eq.
(\ref{bbais}).

Once a Lagrangian formulation has been obtained, we can follow the standard
procedure to quantize this system. Taking $A=0$ as the gauge choice and
introducing ghosts $c\in\P_{>0}\cg$ and anti-ghosts $b\in\P_{<0}\cg$, we get
the gauge
fixed action:
\bea
{\cal S}_{\rm gf}=\k S^-[g]
+ \frac{\k}{4\p x}\int str [\t,e_=]\bdel\t
+\frac{1}{2\p x}\int str\, b\bdel c,
\label{actiongf}
\eea
and the nilpotent BRST charge $\cq_{HR}$:
\bea
\cq_{HR}=\frac{1}{4\p i x}\oint str\left\{ c \left( J -\frac \k 2
e_=-\frac \k 2[e_=,\t]+ \frac 1 2 J^{\rm gh}\right) \right\}.\label{brsch1}
\eea
where $J^{\rm gh}=\frac 1 2 \{b,c\}$. The generators of the extended conformal
algebra are the generators of the cohomology of
$\cq_{HR}$ computed on the algebra $\ca$ which is generated by
$\{b,\hat{J},\t,c\}$, with $\hat{J}=J+J^{\rm
gh}$, and
consisting of all regularized
products of the generating fields and their
derivatives modulo the usual relations
between different orderings, derivatives, etc. The calculation of this
cohomology was performed in \cite{wt,dbt} heavily using spectral sequence
techniques. We summarize the main results.
\begin{enumerate}
\item The cohomology is only non-trivial at ghostnumber 0.
\item For every $sl(2)$ irrep in the decomposition eq. (\ref{decomp}), we
obtain a conformal current $T^{j,\a_j}$, $\a_j\in\{1,\cdots,n_j\}$. They have
the
form:
\bea
T^{j,\a_j}=\sum_{n=0}^{2j}T_{j-\frac n 2}^{j,\a_j}
\eea
where each term in the sum has definite grading $j-n/2$ ($\tau$ has been
assigned grading $0$) and
the leading term $T^{j,\a_j}_j$ is proportional to the appropriate element of
$\P_{\ker ad e_\pp}(\hat{J} +(\kappa/4)
[{\tau},[e_=,{\tau}]])$. The other terms are recursively determined from this.
\item The Virasoro subalgebra is associated to the embedded $sl(2)$
itself. The other currents $T^{j,\alpha_j }$ are primary fields of conformal
dimension $j+1$.
\item The central charge of the system is given by
\bea
c=\frac 1 2 c_{\rm crit} - \frac{(d_B-d_F)\tilde{h}}{\k+\tilde{h}} - 6 y (\k
+\tilde{h}),\label{cpretty}
\eea
where $c_{\rm crit}$ is the critical central charge and $y$ the index of
embedding.
\item The map $T^{j,\alpha_j }\rightarrow T^{j,\alpha_j }_0$ is an algebra
isomorphism. It is the quantum Miura transformation.
\end{enumerate}

Concluding, we see that we obtained a very systematic way to study extended
conformal symmetries. Though not all extensions of the Virasoro algebra
can be obtained this way, a very large class is covered. We end this
section with an example.

The super Lie
algebra $sl(2|1)$ is generated by a bosonic $su(2)\oplus u(1)$ sector: $\{
e_\pp\, ,e_=,e_0,u_0\}$
with $[e_0,e_\pp\, ]=+2e_\pp$, $[e_0,e_=]=-2e_=$ and $[e_\pp\, ,e_=]=e_0$. The
fermionic generators,
$g_\pm$, $\bar{g}_\pm$ are $sl(2)$ doublets, while $g_\pm$ ($\bar{g}_\pm$) have
eigenvalue $+1$ ($-1$) under
$\mbox{ad}_{u_0}$. The remaining commutation relations are easily derived from
the $3\times 3$ matrix
representation
$e_\pp=E_{12}$, $e_==E_{21}$, $e_0=E_{11}-E_{22}$,
$u_0=-E_{11}-E_{22}-2E_{33}$,
$g_+=E_{13}$, $g_-=E_{23}$, $\bar{g}_+=E_{32}$ and $\bar{g}_-=E_{31}$.
The index of the fundamental representation is $x=1/2$.
Only one non-trivial $sl(2)$ embedding is possible.
The WZW model, with action $\k S^- [g]$ on $sl(2|1)$ gives rise to affine
currents
$J=E^\pp e_\pp + E^0 e_0 + E^=e_= + U^0 u_0 + F^+ g_+ +F^- g_- +\bar{F}^+
\bar{g}_++\bar{F}^- \bar{g}_-$.

We now follow the general strategy outlined above. Using the canonical grading
induced by the $sl(2)$
embedding we get for the affine currents:\\[.1cm]

\vspace{-.3cm}
\noindent
\begin{center}
{ \begin{tabular}{|c||c|c|c||c||c|c||c|c||}\hline $
$&$E^\pp$&$E^0$&$E^=$&$U^0$&$F^{+}$&
$F^{-}$&$\bar{F}^{+}$&$ \bar{F}^{-}$ \\ \hline
${\it grade}$&$1$&$0$&$-1$&$0$&$1/2$&$-1/2$&$1/2$&$-1/2$\\ \hline
\end{tabular}}\ .\\[1cm]
\end{center}

\vspace{-.5cm}
\noindent
The gauge group is generated by $\{e_\pp,g_+,\bar{g}_+\}$. Taking $A=A^\pp
e_\pp+A^+g_++\bar{A}^+\bar{g}_+$,
and ${\tau}=\t
g_++\bar{\t}\bar{g}_+$, we get from eq. (\ref{actionst}) the invariant action.
The
quantization yields 4 generators for the BRST cohomology:
\bea
T_{N=2}&=& \frac{2\k}{\k+1}\left(\hat{E}^\pp+
\hat{\bar{F}}^+\t+\hat{F}^+\bar{\t}
-\frac 2 \k \hat{U}^0 \hat{U}^0 +\frac 2 \k \hat{E}^0 \hat{E}^0\right.\nonu &&
\left. -\frac{\k+1}{\k}\del \hat{E}^0-\frac{\k+1}{4} \left( \t\del\bar{\t}
-\del\t \bar{\t}\right)\right)\nonu
G_+&=& \sqrt{\frac{4\k}{\k+1}}\left(\hat{F}^+-\t
\left(\hat{E}^0+\hat{U}^0\right)
+\frac{\k+1}{2}\del\t\right) \nonu
G_-&=&\sqrt{\frac{4\k}{\k+1}}\left(\hat{\bar{F}}^+
-\bar{\t }\left(\hat{E}^0 -\hat{U}^0\right)+\frac{\k+1}{2}\del\bar{\t}\right)
\nonu
U&=&-4\left( \hat{U}^0-\frac \k 4 \t\bar{\t}\right) , \eea
which satisfies the $N=2$ superconformal algebra with $c_{N=2}= -3(1+2\k)$
(this
follows from eq. (\ref{cpretty}), with $y=1$, $\tilde{h}=1$ and $c_{crit}=6$).
The Miura transform yields
the standard
free field realization of the $N=2$ superconformal algebra:
\bea
T_{N=2}&=& \del\vf\del\bar{\vf}-\frac{\sqrt{\k+1}}{2}\del^2(\vf+\bar{\vf})
-\frac 1 2 (\j\del\bar{\j}-\del\j \bar{\j}),\nonu
G_+&=&- \j\del\vf+\sqrt{\k+1}\del\j,\qquad
G_-=- \bar{\j}\del\vf+\sqrt{\k+1}\del\bar{\j} \nonu
U&=&\j\bar{\j}-\sqrt{\k+1}(\del\vf-\del\bar{\varphi}),
\eea
where $\del\vf(z_1)\del\bar{\vf}(z_2)=z_{12}^{-2}$,
$\j(z_1)\bar{\j}(z_2)=z_{12}^{-1}$ and we
introduced some simplifying rescalings:
$\del\vf=2(\k+1)^{-1/2}(\hat{E}^0+\hat{U}^0)$, $\j=\sqrt{\k}\t$,
$\del\bar{\vf}=2(\k+1)^{-1/2}(\hat{E}^0-
\hat{U}^0)$ and  $\bar{\j}=\sqrt{\k}\bar{\t}$.

\setcounter{equation}{0}
\section{Effective actions for 2D gravity}
\noindent
Consider a cft whose fields we collectively denote by $\phi$ and which is
described by an action ${\cal S}[\phi]$. The model has an extended
conformal algebra generated by the currents $T_i[\phi]$, with
$i\in\{1,\cdots, N\}$. The induced action for the extended gravity theory
is, in the light-cone gauge, given by
\begin{eqnarray}
e^{-\Gamma [\mu]}=\int [d\phi]e^{-\Gamma[\phi]-\frac 1 \pi\sum_{i=1}^N\int
\mu^i
T_i[\phi]},
\end{eqnarray}
where the $\mu ^i$ are the generalized Beltrami differentials. The effective
action for the $2D$ extended gravity theory is then
\begin{eqnarray}
e^{-W[\check{T}]}= \int [d\mu ]e^{-\Gamma[\mu ]+\frac 1 \pi\sum_{i=1}^N\int
\mu^i \check{T}_i}.
\end{eqnarray}
Reversing the order of integration, we obtain
\begin{eqnarray}
e^{-W [\check{T}]}= \int
[d\phi]\prod_{i=1}^N\delta(\check{T}-T[\phi])e^{-\Gamma[\phi] },
\end{eqnarray}
which is extremely hard to compute due to the non-trivial
Jacobian.

However, in previous section, we considered a particular form for  the cft
described by
${\cal S}[\phi]$ which will allow for the computation of the Jacobian
\cite{wt,wtk}.
We get for the effective action:
\bea
e^{ -W[\check{T}]}&=&
\int [\d g g^{-1}][d\t][d A][d\m ]
\left(
\mbox{Vol}\left( \P_{>0}{\cal G} \right) \right)^{-1}\nonu
&&\exp \left(-
{\cal S}
-\frac{1}{\p}\sum_{j\in\frac 1 2 {\bf N}}\sum_{\alpha_j=1}^{n_j} \int
\mu_{j,\alpha_j}
\left(T^{j,\alpha_j}-\check{T}^{j,\alpha_j}\right)
\right) ,\label{okok2}
\eea
where ${\cal S}$ was given in eq. (\ref{actionst}). In order to compute the
effective action, we choose the highest weight gauge:
$\t=\P_{> 0}[e_\pp,J_z]=0$ and we find using
$[\d g g^{-1}]=[dJ]\exp \left(
-2\tilde{h} S^-[g]\right)$
\bea
W[\check{T}]=\k_c S_-[g']\label{endresult}
\eea
where $\k_c=\k+2\tilde{h}$ and\footnote{This formula has to be slightly
generalized in
the case that the embedded $sl(2)$ has a non-trivial centralizer in ${\cal G}$
\cite{wt}.}
$\partial g'g'^{-1}=e_=+\sum_{j,\alpha_j}C(j,\alpha_j)
\check{T}^{j,\alpha_j}t_{j,\alpha_j}$.
{}From eq. (\ref{cpretty}) we get the level as a function of the central
charge:
\bea
12 y \k_c=12 y\tilde{h}-\left(c-\frac 1 2 c_{\rm crit}\right)-
\sqrt{\left(c-\frac 1 2 c_{\rm crit}\right)^2- 24 (d_B-d_F)
\tilde{h}y},\label{vv2}
\eea
which provides an all-order expression for the
coupling constant renormalization.
The wavefunction renormalization constants
$C(j,\alpha_j)$ can, albeit using a mild assumption, be computed as well
\cite{wtk}.

\setcounter{equation}{0}
\section{The BRST structure of strings from gauged WZW models}
\noindent
Consider a bosonic string consisting of a matter, a gravity or Liouville and a
ghost sector. The matter
sector is represented by
its energy-momentum tensor $T_m$ which generates the Virasoro algebra with
central charge $c_m$.
The gravity sector is realized in terms of a Liouville field $\vf_L$,
$\del\vf_L(z_1)\del\vf_L(z_2)=-z_{12}^{-2}$,
with energy-momentum tensor $T_L$:
\bea
T_L=-\frac 1 2 \del \vf_L\del
\vf_L+\sqrt{\frac{25-c_m}{12}}\del^2\vf_L, \eea
which has central charge $c_L=26-c_m$. The energy-momentum tensor for the
ghosts,
$T_{gh}=-2b\del c-(\del b) c$,
has central extension $c_{gh}=-26$. The total energy-momentum tensor $T =
T_m + T_L + T_{gh}$
has central charge 0. The BRST current
\bea
J_{BRST}=c\left( T_m+T_L+\frac 1 2 T_{gh}\right)+\a \del \left(c\del
\vf_L\right) + \b \del^2 c,
\label{JBRS}
\eea
with
\bea
\a=-\frac{\sqrt{3}}{6}\left(\sqrt{1-c_m}+\sqrt{25-c_m}\right), \qquad
\b=\frac 1 2 (1-\a^2),
\eea
has only regular terms in its OPE with itself.
The total derivative terms in Eq. (\ref{JBRS}), wich have no influence on the
BRST operator, have precisely been added to
achieve this \cite{bea}. Calling
$G_+\equiv J_{BRST}$ and $G_-\equiv b$, one finds that the current algebra
generated by $T$, $G_+$
and $G_-$ closes, provided a $U(1)$ current $U$ is introduced: \bea
&&T(z_1)T(z_2)=2z_{12}^{-2}T(z_2)+z_{12}^{-1}\del T(z_2),\quad
T(z_1)G_+(z_2)=z_{12}^{-2}G_+(z_2)+z_{12}^{-1}\del G_+(z_2),\nonu
&&T(z_1)G_-(z_2)=2z_{12}^{-2}G_-(z_2)+z_{12}^{-1}\del G_-(z_2),\nonu
&&T(z_1)U(z_2)=-\frac{c_{N=2}}{3} z_{12}^{-3} + z_{12}^{-2} U(z_2)+
z_{12}^{-1}\del U(z_2),\nonu
&&G_+(z_1)G_-(z_2)= \frac{c_{N=2}}{3}z_{12}^{-3}+z_{12}^{-2}U(z_2) +
z_{12}^{-1} T(z_2),\nonu
&&U(z_1)G_\pm (z_2)=\pm z_{12}^{-1}G_\pm (z_2), \qquad
U(z_1)U(z_2)=\frac{c_{N=2}}{3}z_{12}^{-2}, \label{twn2}\eea
where $U\equiv -bc-\a\del\vf_L$ is a modified ghost number current.
Upon untwisting $T_{N=2}=T-\frac 1 2 \del U$, one gets the $N=2$
superconformal
algebra with central extension $c_{N=2}=6\b$.

We now turn to the Hamiltonian reduction and show how to obtain the above from
it.
We want to identify
$G_+$ with the BRST current and $G_-$ with the Virasoro anti-ghost, so a
single
field instead of a
composite.
To achieve this, we have to consider a different grading, according to the
eigenvalues of
$ \frac 1 2 \mbox{ad}_{e_0}+ \mbox{ad}_{u_0}$. We obtain for the gradings of
the
various
currents:\\[.1cm] \begin{center}
{ \begin{tabular}{|c||c|c|c||c||c|c||c|c||}\hline $
$&$E^\pp$&$E^0$&$E^=$&$U^0$&$F^{+}$
&$F^{-}$&$\bar{F}^{+}$&$ \bar{F}^{-}$ \\ \hline
${\it grade}$&$1$&$0$&$-1$&$0$&$3/2$&$1/2$&$-1/2$&$-3/2$\\ \hline
\end{tabular}}\ .\\[1cm]
\end{center}
We follow  the same procedure as before, but whenever we refer to the
grading on the Lie algebra,
we always imply it to be the grading induced by $\frac 1 2 \mbox{ad}_{e_0}+
\mbox{ad}_{u_0}$.
Again we constrain the strictly negatively graded part of the algebra:
\bea
\Pi_{<0} J =\frac \k 2\left(e_= + \psi \bar{g}_+ \right). \eea
The
current $\bar{F}^+$ is a
highest $sl(2)$ weight and will become the leading term of a conformal
current.
But on the same token, $\bar{F}^+$ has a negative grading, so it has to be
constrained. Thus we need
to constrain it in a non-singular way, {\it i.e.} by putting it equal to an
auxiliary field which is inert under
the gauge transformations.
The action which reproduces the constraints is easily obtained: \bea
{\cal S}=\k S^-[g]+\frac{1}{\pi x}\int str\, A(J-\frac \k 2 e_=-\frac \k 2
\Psi
)
+\frac{\k}{2\p x}\int str \Psi \bdel\bar{\Psi}, \eea
where
\bea
A=A^\pp e_\pp+A^+g_++A^-g_-,\qquad
\Psi=\psi\bar{g}_+,\qquad
\bar{\Psi}=\bar{\psi}g_-.
\eea
The gauge invariance is parametrized by $\h\in\P_{>0} sl(2|1)$ or
$\h=\h^\pp e_\pp
+\h^+g_++\h^-g_-$:
\bea
\delta g=\eta g,\quad
\delta A = \bdel \eta + [\eta,A],\quad \delta \bar{\Psi}= \h^-g_-.
\eea
The combined requirements of gauge
invariance and the existence of
a non-degenerate highest weight gauge, requires the introduction of the
field $\bar{\psi}$ conjugate
to $\psi$.
As before, the gauge choice is $A=0$. Introducing ghosts $c = c^\pp
e_\pp+\g^+g_++\g^-g_-
\in \P_{>0} sl(2|1)$ and anti-ghosts $b = b^= e_= + \bar{\b}^+ \bar{g}_+
+\bar{\b}^- \bar{g}_-
\in \P_{<0} sl(2|1)$, we get the gauge fixed action: \bea
{\cal S}_{gf}=\k S^- [g] +\frac{\k}{2\p x}\int str \Psi \bdel\bar{\Psi}
+\frac{1}{2\p x}\int str b\bdel c, \eea
and the nilpotent BRST charge
\bea
Q_{HR}=\frac{1}{4\p i x}\oint str \left\{ c\left( J-\frac \k 2 e_= -\frac \k 2
\Psi + \frac 1 2 J_{gh} \right)
\right\}. \eea
Again, the affine symmetry of the WZW model breaks down to an $N=2$
superconformal symmetry
whose generators the generators of the cohomology ${\cal H}^*({\cal
A},Q_{HR})$,
where $\ca$ is
the algebra generated by $\{ b, \hat{J}=J+ J_{gh}, \psi, \bar{\psi}, c\}$.
We are now in the position to use
spectral sequence techniques, to compute the cohomology. One arrives at
\bea
T_{N=2}&=& \frac{2\k}{\k+1}\left(\hat{E}^\pp+ \psi \hat{{F}}^- -\frac 2 \k
\hat{U}^0 \hat{U}^0
+\frac 2 \k \hat{E}^0 \hat{E}^0\right.\nonu
&&\left. -\del \hat{E}^0-\frac{1}{\k}\del \hat{U}^0 -\frac{\k +1 }{4}\left(
3\psi\del\bar{\psi}
+\del\psi \bar{\psi}\right)\right)\nonu
G_+&=& \frac{2\k^2}{1+\k}\left( \hat{F}^+ + \hat{E}^\pp\bar{\psi}- \frac 2 \k
(
\hat{E}^0
+ \hat{U}^0)\hat{F}^- + \hat{F}^-\bar{\psi} \psi + \del \hat{F}^- \right.\nonu
&&\left.
-\frac{(\k+1)(2\k+1)}{4\k}\del^2 \bar{\psi} + \frac 2 \k ( \hat{E}^0\hat{E}^0
- \hat{U}^0\hat{U}^0 )
\bar{\psi}\right.\nonu &&\left. - \del (\hat{E}^0-\hat{U}^0) \bar{\psi}
-\frac{1+\k}{2}\psi \del \bar{\psi}
\bar{\psi} + \frac{2(1+\k)}{\k} \del \bar{\psi}\hat{U}^0\right) \nonu
G_-&=&\psi\qquad\qquad
U=-4\left( \hat{U}^0+\frac \k 4 \psi\bar{\psi}\right) , \label{rrr}\eea
which satisfies the $N=2$ superconformal algebra with $c_{N=2}= -3(1+2\k)$.
The Miura transform is again given by the algebra isomorphism which maps the
currents in eq. (\ref{rrr}) on their grade 0 part (the $\Psi$ and $\bar\Psi$
fields have grade 0). This together with the OPE's
$\hat{E}_0(z_1)\hat{E}_0(z_2)=- \hat{U}_0(z_1)\hat{U}_0(z_2) = (\k + 1)/8 \,
z_{12}^{-2}$
and $\psi(z_1) \bar{\psi}(z_2) = 1/\k \, z^{-1}_{12}$ gives the desired
realization
of the $N=2$ algebra.
Indeed, identifying $b \equiv \psi$, $c \equiv \k \bar{\psi}$, $\del \varphi_L
\equiv \sqrt{8 / (\k +1)} \hat{U}_0$
and $\del \varphi_m \equiv i\sqrt{8 / (\k +1)} \hat{E}_0$, and
\bea
T_m=-\frac 1 2 \del\vf_m\del\vf_m + i \frac{\k}{\sqrt{2(\k+1)}}\del^2\vf_m,
\eea
precisely reproduces, upon twisting, the non-critical string theory discussed
at the beginning of this section
with
\bea
c_m=1-6\left (\sqrt{\k+1}-\frac{1}{\sqrt{\k+1}}
\right)^2.
\eea

This program has been carried out in general \cite{usnew}. The classification
of all
possible $sl(2|1)$ embeddings in super Lie algebras, yields all possible
extensions of the $N=2$ superconformal algebra. The subset of
embeddings of $sl(2|1)$ in ${\cal G}$ which allow for a stringy interpretation,
are those
for which the adjoint representation of ${\cal G}$ decomposes into $sl(2|1)$
irreps $(b,j)$,
with $b=0$
\footnote{This requirement follows from considering the ghost number
assignments in the resulting string theory. Typical $sl(2|1)$ irreps $(b,j)$,
$b\neq \pm j$, consist of 4 sl(2) irreps: $|j,b>$, $|j-1/2,b\pm 1/2>$ and
$|j-1,b>$, where the second label denotes half the $u(1)$ eigenvalue. Atypical
irreps $(\pm j,j)$ contain only 2 $sl(2)$ irreps. In the string theory, the
$u(1)$ eigenvalue gets identified with the ghostnumber.}.
This is only possible for the following cases:
\begin{enumerate}
\item $sl(m|n)$\\
A principal embedding of $sl(2|1)$ in $p\, sl(2j+1|2j)\oplus q\, sl(2j|2j+1)$,
which in its turn is regularly
embedded in $sl(m|n)$ with $p,\, q\geq 0$, $j\in\frac 1 2 {\bf N}$,
$m=p(2j+1)+2qj$ and $n=2pj+q(2j+1)$.
\item $osp(m|2n)$\\
The regular embedding of $osp(2|2)$  in
$osp(m|2n)$.
\item $D(2,1,\a)$\\
$osp(2|2)$ as a regular subalgebra of $D(2,1,\a)$.
\end{enumerate}
After the reduction we are left with the $N=2$ superconformal currents together
with a set of $N=2$ multiplets each of  which generically contains four
currents which yields some extension of the $N=2$ superconformal algebra. The
currents fall
into unconstrained $N=2$
multiplets each containing four currents, say $Y(x)$, $H_+(x)$, $H_-(x)$ and
$Z(x)$ of conformal dimensions
$h+1$, $h+1/2$, $h+1/2$ and $h$. Twisting
amounts to replacing $Y(x)$ by $X(x)\equiv Y(x)+\frac 1 2 \del Z(x)$.
If $G_-$ and all fields of the $H_-$ type are realized as single fields,
something which is achieved by modifying the canonical  $sl(2)$ grading
by adding a multiple
of the $u(1)$ charge to it,
then we can view the system as a string theory. Indeed, one gets
from $N=2$ representation theory the following results,
\bea
Q=\frac{1}{2\p i}\oint dz\, G_+(z),
\eea
is the BRST charge of the string, satisfying $Q^2=0$. The $G_-$ current and all
currents of
the $H_-$ type are the anti-ghosts. The total symmetry currents (matter +
gravity + ghosts) are the energy-momentum tensor $T=T_{N=2}+\frac 1 2 \del U$
and the currents of the $X$ type and
they are indeed the BRST transform of the corresponding antighosts:
\bea
T=[Q,G_-]_+\qquad\quad X=[Q,H_-]_\pm.
\eea
The explicit construction of the gauge invariant action and its quantization
has been
performed in
\cite{usnew}. Modulo some complications mostly of a technical nature, the
general discussion parallels the example given at the beginning of this
section. The stringtheories which can be obtained this way are:
\begin{enumerate}
\item for $sl(2|1)\rightarrow p\,sl(2j+1|2j)\oplus q\,
sl(2j|2j+1)\hookrightarrow
sl(p(2j+1)+2qj|2pj+q(2j+1))$\\
one has that the matter sector of the string theory corresponds to
the reduction:
\bea
sl(2)\rightarrow p\, sl(2j+1)\oplus q\, sl(2j+1)\hookrightarrow
sl(p(2j+1)|q(2j+1)).
\eea
\item for $osp(2|2)\hookrightarrow osp(m|2n)$\\
the matter sector is now given by the reduction
\bea
sp(2)\hookrightarrow sp(2n) \hookrightarrow osp(m-2|2n).
\eea
\end{enumerate}
In particular, we get that the first case covers $W_n$ strings by
choosing $j=(1/2)(n-1)$, $p=1$ and $q=0$ \cite{BLNW}
and by choosing $m=N+2$ and $n=1$ in
the second case, we get the $N$-extended superstrings \cite{BLLS}.
At this point it is
an interesting open question to find out which string theory
corresponds to the reduction of $D(2,1,\alpha )$. A priori, one would
expect the $N=2$ superstring. However, the BRST structure of $N=2$
superstrings was explicitely studied in \cite{BLLS} where it was found
that $N=2$ superstrings were covered by $osp(4|2)$ or in other words
$D(2,1, \alpha =1)$.

\setcounter{equation}{0}
\section{Discussion}
\noindent
We saw that embeddings of $sl(2)$ in (super) Lie algebras provide a
powerful method to construct cft's and $2D$ gravity theories. The
embedding completely determines the conformal symmetry. The precise
realization of this symmetry depends on the grading, or equivalently the gauge
group,
chosen on the Lie
algebra. Through the study of $sl(2|1)$ embeddings, one not only classifies
extended $N=2$ superconformal symmetries but using
a particular grading , one
gets the explicit
construction of string theories as well. The last result follows from the
observation that the BRST strucure of a string theory is fully
characterized by a twisted extended $N=2$ superconformal symmetry.

Several open problems remain. Most of the work performed till now was done
at the level of constructing actions, symmetry currents, BRST charges, ...
Though it is a non-trivial result that these things can be obtained in an
algebraic, almost algorithmic way ({\it e.g.} before the development of
these methods, the only way to construct the BRST charge of strings was
through trial and error, a method which in most cases led to formidable
computer calculations), one would like to push this program further and
obtain results on spectrum, correlation functions, ... Due to recent
results in \cite{driel} where the structure of singular and subsingular vectors
in affine Lie algebra Verma modules got unravelled, one hopes that at
least partition functions become calculable.

\vspace{1cm}

\noindent {\bf Acknowledgments} I'd like to thank A. Boresh, K. Landsteiner, W.
Lerche,
E. Ragoucy, P. Sorba, K. Thielemans and W. Troost for pleasant collaborations
which led to the results reviewed here.

\end{document}